\documentclass{iopart}

\usepackage{ifthen}
\usepackage{ifpdf}
\usepackage{color}

\ifpdf
\usepackage{graphicx}
\usepackage{epstopdf}
\else
\usepackage{graphicx}
\usepackage{epsfig}
\fi
\usepackage{latexsym}
\usepackage{amssymb}
\usepackage{bm}
\newcommand{\const}{\mbox{const}}

\newcommand{\mylabel}[1]{\label{#1}} 
\newcommand{\beq}{\begin{eqnarray}}
\newcommand{\eeq}{\end{eqnarray}} 
\newcommand{\be}[1]{\begin{eqnarray}\ifthenelse{#1=-1}
{\nonumber}{\ifthenelse{#1=0}{}{\mylabel{e#1}}}}
\newcommand{\ee}{\end{eqnarray}} 

\newcommand{\Eq}[1]{\textcolor{blue}{Eq.~(\ref{#1})}}
\newcommand{\Fig}[1]{\textcolor{blue}{Fig.}~\ref{#1}}
\newcommand{\hide}[1]{[hidden text]}

\newcommand{\ei}{\hat{a}}
\newcommand{\eidag}{\hat{a}^{\dag}}
\newcommand{\hn}{\hat{n}}

\newcommand{\Jx}{\hat{J}_x}
\newcommand{\Jy}{\hat{J}_y}
\newcommand{\Jz}{\hat{J}_z}

\begin{document}

\title[phase space tomography]{Temporal fluctuations in the bosonic Josephson junction as a probe for phase space tomography}

\author{Christine Khripkov$^1$, Doron Cohen$^2$, and Amichay Vardi$^1$}

\address{Departments of $^1$Chemistry and $^2$Physics, Ben-Gurion University of the Negev, Beer-Sheva 84105, Israel}

%\author{A. Vardi}
%\affiliation{Department of Chemistry, Ben-Gurion University of the Negev, P.O.B. 653, Beer-Sheva 84105, Israel}
%\author{D. Cohen}
%\affiliation{Department of Physics, Ben-Gurion University of the Negev, P.O.B. 653, Beer-Sheva 84105, Israel}

\begin{abstract}
We study the long time dynamics of the reduced one-particle Bloch-vector ${\bf S}$ of a two-mode Bose-Hubbard model in the Josephson interaction regime, as a function of the relative phase and occupation imbalance of an arbitrary coherent preparation.  We find that the variance of  the long time fluctuations of ${\bf S}$ can be factorized as a product of the inverse participation number $1/M$ that depends only on the preparation, and a semi-classical function $C(E)$ that reflects the phase space characteristics of the pertinent observable.  Temporal fluctuations can thus be used as a sensitive probe for phase space tomography of quantum many-body states.
\end{abstract} 

%\pacs{03.65.Xp,03.75.Mn,42.50.Xa}

%%%%%%%%%%%%%%%%%%%%%%%%%%%%%%%%%%%%%%%%%%%%%%%%%%%%%%%%%%%%%%%%%%%%%%%%%%%%%%%%%%%%%%%%%%%%%%%%%%%%%%%

The two mode Bose-Hubbard Hamiltonian (BHH) appears in different guises in a perplexing variety of fields. Cast in spin form, it is known in nuclear physics as the Lipkin-Meshkov-Glick (LMG) model of shape phase transitions \cite{LMG}.  It is broadly used to describe interacting  spin systems \cite{Botet83} and serves as a paradigm for squeezing and entanglement \cite{SpinSqueezing}. As such, it offers schemes for the generation of squeezed states for optical interferometry below the standard quantum limit \cite{OpticalInterferometry},  and its matter-wave equivalent \cite{AtomInterferometry}. It is commonly employed to describe the Josephson dynamics in systems of bosonic atoms in double-well potentials \cite{DoubleWell}  and suggests prospects for the generation of macroscopic superposition states \cite{Cat}. The same model is also known in condensed matter physics as the integrable dimer model \cite{Dimer} with applications to the dynamics of small molecules, molecular crystals, and self-trapping in amorphous semiconductors.

Like the paradigmatic Jaynes-Cummings model in quantum optics \cite{JC}, the bimodal BHH dynamics with a coherent spin state preparation exhibits a series of collapses and revivals of its single-particle coherence due to interactions \cite{Revivals,Boukobza09,Chuchem10,Egorov11}. These recurrences are manifested in the collapse and revival of the Rabi-Josephson population oscillations, or of the multi-realization fringe visibility, when the two condensates are released and allowed to interfere.  Below we study the long time BHH dynamics for general coherent spin preparations $|\theta,\phi\rangle$. In such states all particles occupy a single superposition of the two modes, with a normalized population imbalance $S_z=\cos(\theta)$ and a relative phase~$\phi$. 

The characteristics of the two-mode BHH dynamics strongly depends on the dimensionless interaction parameter
\be{1001}
u \ \ = \ \ UN/K~,
\eeq
where $U$ is the interaction strength, $N$ is the total particle number, and $K$ is the hopping amplitude. In the linear Rabi regime (${|u|<1}$) time evolution is straightforward because the interaction is weak and the nature of the dynamics is essentially single-particle. Accordingly, one observes only coherent Rabi oscillations in the population difference with a typical frequency
\be{1002}
\omega_J \ \ \equiv \ \ \sqrt{K(K+UN)} \ \ = \ \ K\sqrt{1+u}~, 
\eeq
which reflects mainly the coupling~$K$ between the two modes,  
accompanied by a slow loss of single particle coherence.

The dynamics in the highly nonlinear Fock regime ($|u|>N^2$) are also fairly simple because the evolution reflects the Fock basis expansion of the initial coherent preparation. For such strong interactions  the two-mode BHH generates precisely the same dynamics as the many-mode BHH of a BEC in an optical lattice, because the local modes are essentially decoupled, hence the dynamics is fully captured by the Gutzwiller ansatz  of a direct product of single-site states, each of which is a coherent wavepacket of number states \cite{Jaksch98, Orzel01,Greiner02, Will10}.  This allows for monitoring the fringe visibility in single shot interferometery of an optical lattice, rather than repeating a two-mode experiment many times. The expected  recurrences have been observed experimentally for optical lattices with relatively small occupation numbers \cite{Greiner02,Will10}  with a striking demonstration of exceptionally long  time dynamics, allowing to probe effective multi-body interactions through the dependence of $U$ on the number of atoms \cite{Johnson09}.

The dynamics in the Josephson regime (${1<|u|<N^2}$) is  by far richer and more intricate, reflecting the coexistence of three distinct phase space regions \cite{Boukobza09,Chuchem10}. Unlike the Fock-space recurrences, which only depend on the population imbalance, the Josephson coherence dynamics is also highly sensitive to the relative phase. 

Previous work has been limited to short-time dynamics of specific preparations that were of contemporary experimental relevance, e.g. small perturbation of the ground state that results in Josephson oscillations, or a large population imbalance that leads to self-trapping \cite{Smerzi97}. 
Here we greatly expand this scope of view and find a fundamental relation which accurately predicts the fluctuations of any observable over much longer timescales.

We adopt a global, tomographic approach by characterizing the long time temporal quantum fluctuations for {\em all} possible coherent preparations. This appears  to be a formidable task, but as shown below, a relatively simple semi-classical perspective provides an adequate framework for the required analysis.  

Most interestingly, our results show that while long-time fluctuations can not be predicted solely from the semiclassical dynamics (i.e. from the classical propagation of phase-space distributions), they do factorize into a semiclassical term and a quantum term proportional to the number of participating states. We shall discuss how this observation can be utilized in the summary section.

%%%%%%%%%%%%%%%%%%%%
\begin{figure}
\centering
\includegraphics[width=0.8\hsize] {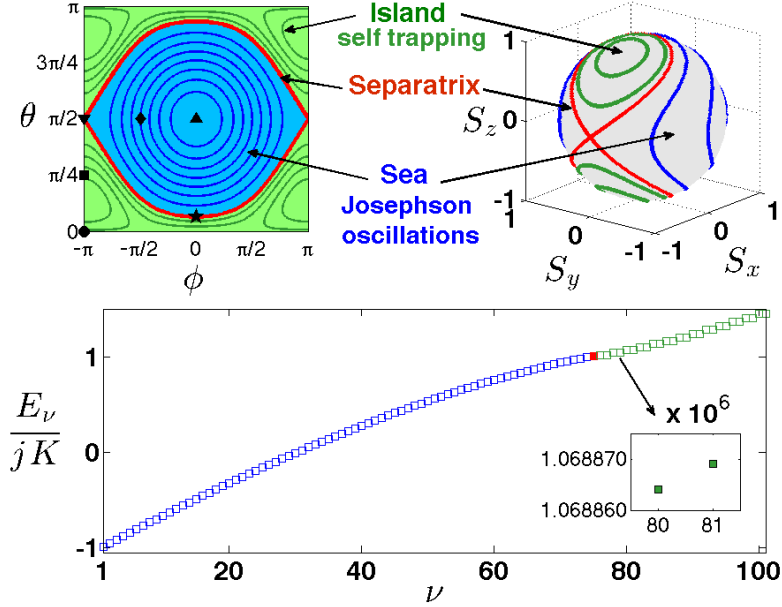}
\caption{(color online)  Phase-space structure (above) of the bosonic Josephson junction in the Josephson regime ($u=2.5$). Lines depict equal energy contours, i.e. classical trajectories. A separatrix trajectory with an isolated hyperbolic point at $(\theta,\phi)=(\pi/2,\pi)$ separates Rabi-Josephson oscillations around the ground state in a $K$-dominated 'sea' from nonlinear self-trapped phase-oscillations in two high-energy, $UN$-dominated  'islands'. Symbols denote the coherent preparations used in \Fig{fig2}. The energy spectrum of the system (below) correspondingly includes a low energy harmonic sea part with characteristic spacing $\omega_J$, a high energy islands part of odd and even macroscopic-cat-like doublets with inter-doublet spacing of $\omega_+$ and intra-doublet tunnel splitting of $\omega_d$ (see insert), and an intermediate separatrix part with spacing $\omega_x$.}
\label{fig1}
\end{figure}  
%%%%%%%%%%%%%%%%%%%%%

%%%%%%%%%%%%%%%%%%%%%%%%%%%%%%%%%%%%%%%%%%%%%%%%%%%%%%%%%%%
\section{The BHH}

We consider a similar scenario to that in Ref.~\cite{Will10}, which observed long time collapses and revivals in the Fock regime.  In the Josephson regime, the dynamics of a lattice with one mode per site is quite different from the two-mode dynamics. However, the two-mode model can be realized with two spin components in each isolated site \cite{Widera08} or with an array of independent double wells \cite{Folling07}, thus retaining the convenience of single-shot measurements. We note recent work on BECs in 1D double-well traps reporting the breakdown of the lowest Bloch band BHH model at interaction parameter values as low as $u=2.15$ for this realization \cite{Sakmann09,Luhmann12}. 

Assuming that no bias field is applied, the pertinent BHH is, 
\beq
\mathcal{H} &=&-\frac{K}{2}\left(\eidag_1\ei_2+\eidag_2\ei_1\right) 
\nonumber \\
&&+\frac{U}{2}\left[\hn_1\left(\hn_1-1\right)+\hn_2\left(\hn_2-1\right)\right],
\label{BHH}
\eeq
where $\ei_i$ and $\eidag_i$  are bosonic annihilation and creation operators, respectively. 
The particle number operator in mode~$i$ is $\hn_{i}=\eidag_{i}\ei_{i}$.  Since the total particle number $\hn_1+\hn_2=N$ is conserved, we can eliminate respective $c$-number terms and obtain the BHH in spin form,
\begin{equation}
\label{SBHH}
\mathcal{H} \ =- K\Jx+ \ U\Jz^2~,
\end{equation}
where $\Jx{=}(\eidag_1 \ei_2{+}\eidag_2\ei_1)/2$, 
and $\Jy{=}(\eidag_1\ei_2{-}\eidag_2\ei_1)/(2i)$, 
and $\Jz{=}(\hn_1 {-} \hn_2)/2$ obey canonical SU(2) commutation relations. Number conservation becomes angular momentum conservation with ${j=N/2}$. Below we assume for simplicity that the interaction is repulsive $U>0$, but the $U<0$ case (to the extent that the particle number is sufficiently small that the attractive BEC is stable) amounts to a simple transformation $K\mapsto-K$, and $E\mapsto -E$. Thus the phase space with attractive interaction is simply an inverted mirror image of the repulsive-interaction case and there is no loss of generality.

%%%%%%%%%%%%%%%%%%%%%%%%%%%%%%%%%%%%%%%%%%%%%%%%%%%%%%%%%%%
\section{The Bloch vector}

In the ``Bloch picture" of the BHH, the reduced one-particle density matrix $\rho^{sp}_{i,j}\equiv\langle \eidag_i\ei_j\rangle$  of each many-body state is represented by the normalized Bloch vector,
\beq
{\bf S} \ \ \equiv \ \ \langle{\bf J}\rangle/j~.
\eeq
The Bloch vector components $S_i$, correspond to the projection of $\rho_{sp}$ onto the Pauli basis $\left\{\sigma_i\right\}_{i=x,y,z}$:
\beq
\rho^{sp}=\frac{1}{2}\left({\bf 1}+S_x\sigma_x+S_y\sigma_y+S_z\sigma_z\right)~.
\label{spdm}
\eeq
The $z$ projection $S_z=\cos(\theta)$ corresponds to the normalized population imbalance, whereas the azimuthal angle $\phi=\arctan(S_y/S_x)$ corresponds to the relative phase between the modes. The components $S_x,S_y$ can be directly found experimentally  by conducting fast $\pi/2$ rotations about $S_y,S_x$ respectively, thus projecting them onto a measurable population imbalance. Alternatively, $\phi$ can be deduced from the position of fringes in an interferometric measurement. The Bloch vector's length $S$ corresponds to the single-particle coherence, which defines the best fringe visibility one may expect  to measure by proper manipulation, i.e. if we are allowed to perform any SU(2) rotation.

%%%%%%%%%%%%%%%%%%%%%%%%%%%%%%%%%%%%%%%%%%%%%%%%%%%%%%%%%%%
\section{Phase space}

The classical phase space structure of the BHH is set by the dimensionless interaction parameter $u$ of \Eq{e1001}.  Its characteristics in the  three interaction regimes are discussed in great detail elsewhere \cite{DoubleWell,Boukobza09,Chuchem10}. 
In \Fig{fig1} we plot the equal-energy contour lines and the pertinent phase-space regions in the Josephson regime (${1<u<N^2}$). Two nonlinear islands are separated from a nearly-linear sea region by a separatrix trajectory.  The sea trajectories correspond to Rabi-Josephson population oscillations around the ground state, whereas the island trajectories correspond to self-trapped phase-oscillations \cite{Smerzi97}. 
In the Fock regime (${u>N^2}$) the sea becomes too small 
to support quantum states, while in the opposite limit - 
in the Rabi regime (${u<1}$) - the islands disappear, 
so that only Rabi-type oscillations are feasible.

%%%%%%%%%%%%%%%%%%%%%%%%%%%%%%%%%%%%%%%%%%%%%%%%%%%%%%%%%%%
\section{Spectrum}

This classical phase-space structure results in a quantum eigenenergy spectrum with three parts (see \Fig{fig1}, bottom panel). For repulsive interaction the lowest part of the spectrum is nearly harmonic, corresponding to the quantization of linear sea trajectories, with characteristic level spacing $\omega_J$. Approaching the separatrix energy, level-spacing becomes smaller due to the nonlinearity, with characteristic spacing of 
\beq
\omega_x \ \ = \ \ [\log(N^2/u)/2]^{-1} \ \omega_J~.
\eeq
The high-energy part of the spectrum consists 
of doublets at $E \approx Um^2$, 
with $2Um$ spacing, approaching the value $\omega_+=UN$ 
as $m\rightarrow j$. These states correspond to macroscopic 
cat-like superpositions of quantized island trajectories \cite{Cat}. 
The internal doublet splitting $\omega_d$
between such odd and even macroscopic cat states, 
reflects the many-body quantum tunneling frequency  
between the islands, and vanishes exponentially 
as the particle number $N$ is increased.   
The quantum tunneling time between the islands 
is thus characteristically many orders of magnitude 
larger than the classical periods associated
with the frequencies $\omega_{J}$, 
and $\omega_{x}$, and $\omega_{+}$.

%%%%%%%%%%%%%%%%%%%%%%%%%%%%%%%%%%%%%%%%%%%%
\section{Evolution} 

We study the dynamics induced by the Hamiltonian of \Eq{SBHH}, starting from an arbitrary spin coherent state preparation,
\begin{eqnarray}
|\theta,\phi\rangle&\equiv&\frac{1}{N!}\left[\cos(\theta/2)\eidag_1+\sin(\theta/2)e^{i\phi}\eidag_2\right]^N |{\rm vac}\rangle\nonumber\\
~&=&\exp({-i\phi\Jz})\exp({-i\theta\Jy}) \ |J_z=j\rangle,
\label{css}
\end{eqnarray}
where $|{\rm vac}\rangle$ and $|J_z=j\rangle$ are the vacuum states of the Heisenberg-Weyl and SU(2) algebras, respectively. The preparation of such arbitrary coherent states can be attained via a two step process as implied 
by \Eq{css} and demonstrated experimentally in Ref. \cite{Zibold10},  in which $\theta$ is set by a coupling pulse and $\phi$ by a bias pulse.

%%%%%%%%%%%%%%%%%%%%%%%
\begin{figure}
\centering
\includegraphics[width=0.8\textwidth] {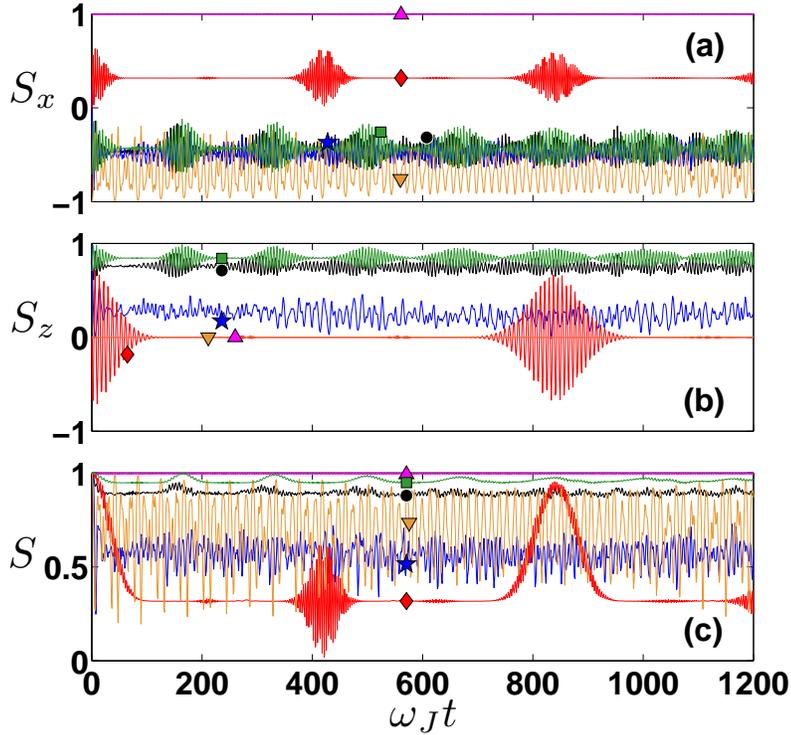}
\caption{(color online) 
Quantum BHH dynamics of the 
Bloch vector components $S_{x,z}$ (a,b) 
and of the one-particle coherence $S$ (c)  
for the representative coherent preparations  marked in \Fig{fig1}. The BHH parameters here and in all subsequent figures are ${N=100}$ and ${u=2.5}$, within the Josephson regime.
}
\label{fig2}
\end{figure}  
%%%%%%%%%%%%%%%%%%%%%%%%

The intricacy of the Josephson regime quantum dynamics is illustrated in \Fig{fig2}, where we plot the time-evolution of $\bf S$, as generated by the BHH \Eq{SBHH}, for several representative coherent preparations (corresponding to the symbols in \Fig{fig1}). 
It is clear that different preparations lead to qualitatively different behavior, depending on the initial population imbalance and on the relative phase.  Moreover, different preparations located on the same classical trajectory produce dramatically different recurrence patterns (see e.g. the differences in the coherence dynamics between the two on-separatrix preparations marked by star and inverted triangle in \Fig{fig1} and \Fig{fig2}c). The cause of this diversity is that different coherent preparations sample different parts of the spectrum. Each coherent spin state constitutes a superposition of eigenstates that can be associated with qualitatively different regions in the corresponding classical phase space. By contrast, in the Rabi and Fock regimes eigenstates occupy, so to say, a single component phase space that allows only one type of motion.

%%%%%%%%%%%%%%%%%%%%%%%%%%%%%%%%%%%%%%%%%%%%
\section{Fluctuations} 

For each preparation $|\theta,\phi\rangle$ we characterize the temporal fluctuations of the expectation values $A(t) = \langle \hat{A} \rangle_t$ of the pertinent observables, by their time-average  
\beq
\overline{A(t)}\equiv\frac{1}{T}\int_0^T A(t)dt~,
\eeq
and by their variance 
\beq
\sigma^2_A \ \ \equiv \ \  \overline{A^2(t)}-{\overline{A(t)}}^2~,  
\eeq
taken over a long time $t<T$ compared to the collapse and revival timescale. We note that $\omega_{J,x,+}^{-1}\ll T\ll \omega_d^{-1}$, i.e. that the averaging time is long with respect to all classical periods but still short with respect to the many-body tunneling time between the islands. The experimental feasibility of such long-time coherence measurements has been demonstrated in Ref.~\cite{Will10} where the fringe-visibility dynamics in a BEC of $^{87}$Rb atoms in an optical-lattice, has been observed over $\sim10$ms duration, compared to $\sim 100 \mu$s collapse-revival period.

Plotting the long time average $\bar{A}$ of 
an observable as a function of $(\theta,\phi)$ we 
obtain an {\em image} of phase space. We refer to 
the information that can be extracted from such an  
image as ``phase-space tomography". Our
main result as outlined below, shows that an image of the variance $\sigma^2_A$ 
provides valuable complementary information that 
is missing in the image of $\bar{A}$.

\begin{figure}
\centering
\includegraphics[width=0.8\hsize] {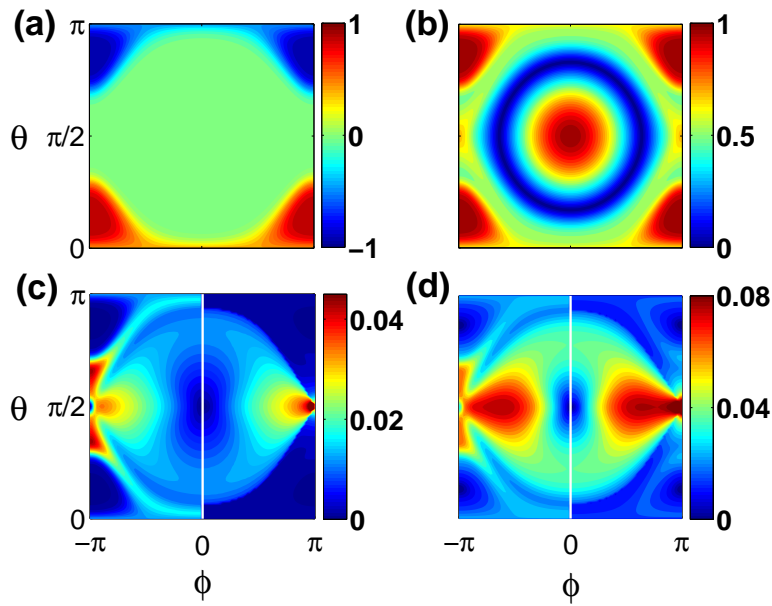}
\caption{(color online) 
(a) Tomographic image of the population imbalance $\overline{S_z}$. 
(b) Tomographic image of $|\overline{{\bf S}}|$.
(c) Tomographic image of $\sigma^2_{S_z}$.
(d) Tomographic image of $\sigma^2_{{\bf S}}$.
The left side of panels c,d correspond to the numerical 
results based on \Eq{dispers}, 
whereas the right side is the factorization of \Eq{factorization}. }
\label{fig3}
\end{figure}

To illustarte this point,  we plot in \Fig{fig3}a,b such images of $\overline{S_z}$ 
and of $|\overline{{\bf S}}|$ for all the possible coherent preparations $|\theta,\phi\rangle$. 
In the lower panels \Fig{fig3}c,d we plot the 
complementary images of $\sigma^2_{S_z}$ and of the vector variance 
$\sigma^2_{{\bf S}}=\sum_{i=x,y,z} \sigma^2_{S_i}$.

The top mean-value images are fairly straightforward to understand in classical terms. 
For example, in the phase space image of ${\overline S_z}$ (\Fig{fig3}a sea trajectories have zero average population imbalance, whereas self-trapped island trajectories retain a finite imbalance.  Note that the formal infinite-time average of the population imbalance is identically zero also for island preparations, due to the definite mode-exchange parity of energy eigenstates.  In other words, a classically self-trapped preparation in one of the islands will eventually quantum-mechanically tunnel to the other island on a $\omega_d^{-1}$ timescale. However, this formal observation is of no physical relevance due to the scaling of $\omega_d$ with $N$. For example, in our simulations here it corresponds to no less than $10^{13}$ Josephson periods before localization is lost. 

It is much more difficult to provide a purely classical interpretation for the bottom fluctuation  images (e.g. the phase space image of $\sigma^2_{S_z}$ in \Fig{fig3}c). 
Naively, we could attempt to attribute them to fluctuations 
in the semiclassical propagation of distributions in phase-space, 
in the spirit of the truncated Wigner approach.
If this was the case, the fluctuations of all coherent preparations along the same classical energy contour would have been the same, because the initial Gaussian distribution would have smeared all over this classical trajectory. The long time averages and fluctuations would have thus corresponded to the phase-space averages and variances along classical trajectories, and accordingly \Fig{fig3}c would reflect, like \Fig{fig3}a, the classical structures of \Fig{fig1}.

However, the imbalance fluctuations (\Fig{fig3}c) show a far more complex pattern which does not seem to be directly related to the mean-field trajectories. Similarly, while the image for the length of the average Bloch vector (\Fig{fig3}b) matches the classical structure, its fluctuations (\Fig{fig3}d) can not be attributed to classical features alone. Below we analyze and explain these observed patterns, showing  that they are the product of the described semiclassical factor, obtained from classical propagation, and a quantum factor which is inversely proportional to the number of eignestates $M$ participating in a given coherent preparation.

%%%%%%%%%%%%%%%%%%
%%%%%%%%%%%%%%%%%%

%%%%%%%%%%%%%%%%%%%%%%%%%%%%%%%%%%%%%%
\section{Analysis} 

In order to deduce the exact time average of any $A(t)$, 
we expand it in the energy basis as 
\beq
\label{Aoft}
A(t) \ \ = \ \ \sum_{\nu,\mu} c_\nu^* c_\mu A_{\nu\mu}\exp[(E_\nu-E_\mu)t/\hbar],
\eeq
where $|E_\nu\rangle$ are the BHH eigenstates,   
$c_\nu=\langle E_\nu|\psi\rangle$ are the 
expansion coefficients of the initial 
state $|\psi\rangle=|\theta,\phi\rangle$, 
and $A_{\nu\mu}=\langle E_\nu | {\hat A} | E_\mu \rangle$. 
The long-time average eliminates the oscillating terms, 
hence 
\beq
\overline{A(t)}=\sum_\nu p_\nu A_{\nu\nu}~, 
\eeq
with probabilities $p_\nu\equiv|c_\nu|^2$, while the variance is 
\begin{equation}
\label{dispers}
\sigma_A^2 \ \ = \ \ \sum_{\nu\neq\mu} p_\nu p_\mu |A_{\nu\mu}|^2~.
\end{equation}
Again we note that extremely
slow tunneling oscillations 
with $E_\nu-E_\mu \sim \omega_d$,
due to the doublets in the spectrum, 
are not eliminated by averaging over 
the experimentally accessible times. 
The doublets can certainly be considered 
degenerate on any realistic timescale.

The matrix elements in \Eq{dispers} can be evaluated semi-classically
using the following prescription~\cite{Cohen01,Chuchem10}:  
a classical trajectory of energy $E$ is generated  
using the BHH mean field equations of motion, 
and $A_{cl}(t)$ is calculated; 
then the classical power-spectrum $\tilde C_A^{cl}(\omega)$ 
is obtained via a Fourier transform of ${[A_{cl}(t)-\overline{A_{cl}}]}$; 
and finally the result is divided by the mean level spacing~$\varrho$ 
at that energy, providing the approximation,
\beq
|A_{\nu\mu}|^2 \ \ \approx \ \ 
\frac{1}{2\pi\varrho} \ 
\tilde C_A^{cl}(E_\nu{-}E_\mu)~. 
\eeq
This is a very general procedure which is usually applied to chaotic systems, but it applies equally well to the integrable non-linear motion of the two-mode BHH.  The number of eigenstates that contribute to \Eq{dispers}, is conventionally evaluated as the {\em participation number} 
\beq
M \ \ \equiv \ \ \frac{1}{\sum_\nu p_\nu^2}~. 
\eeq
Assuming $M\gg1$, approximating $p_\nu\approx 1/M$, and neglecting non-participating eigenstates, we obtain that,
\begin{equation}
\sigma^2_A \ \ =  \ \ \frac{1}{M}C_A(E), 
\label{factorization}
\end{equation}  
where,
\begin{equation}
C_A(E) \ \ = \ \ 
\sum_{|r|>0} |A_{\nu\mu}|^2=\int {\tilde C}_A^{cl}(\omega)\frac{d\omega}{2\pi}.
\label{powerspectrum}
\end{equation}
Above $r=(\nu-\mu)$ is the diagonal coordinate of the matrix, 
and it is implicit that the summation is carried out over 
a section $\nu+\mu= \const$ such that $(E_\nu+E_\mu)/2 \sim E$. 
Note that the time variation of $A_{cl}(t)$
is non-linear but periodic, accordingly
the integral in \Eq{powerspectrum} is related, up to
a form factor, to the classical amplitude.

\Eq{factorization} with the definition \Eq{powerspectrum} constitutes
our main result. Given any observable with a fluctuating expectation value, 
that has long time average $\overline{A}$, 
its variance $\sigma^2_A$ can be approximated 
accurately as a product of the quantum term $1/M$, 
and a semi-classical term $C_A(E)$.
The latter corresponds to the classical fluctuations of~$A$ 
along a mean-field trajectory that has an energy~$E$. 
Below we show that indeed this factorization results 
in the apparently complex patterns of \Fig{fig3}c,d.

%%%%%%%%%%%%%%%%%%%
\begin{figure}
\centering
\includegraphics[width=0.8\hsize] {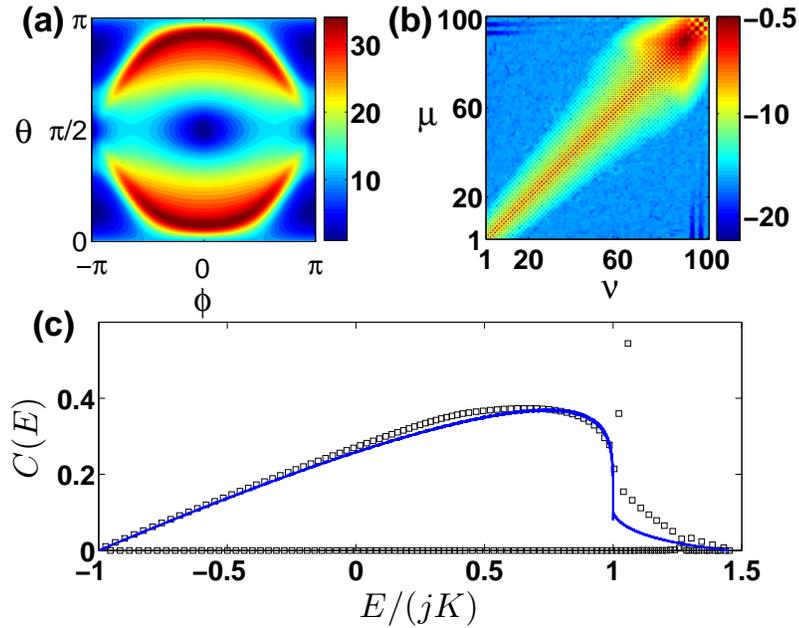}
\caption{(color online)  (a) The participation number $M(\theta,\phi)$ for all coherent preparations $|\theta,\phi\rangle$ (b) image of the matrix elements $|A_{nm}|$ for the population imbalance $A=J_z$, with color-scale in log10 units; (c) The power spectrum $C_A(E)$ for the same observale, evaluated according to the middle (symbols) and the r.h.s. (lines) 
of \Eq{powerspectrum}. The separatrix energy is $E/(jK)=1$.}
\label{fig4}
\end{figure}  
%%%%%%%%%%%%%%%%%%%

%%%%%%%%%%%%%%%%%%%%%%%%%%%%%%%%%
\section{Numerical verification} 

The required ingredients for the calculation of the variance $\sigma^2_A(\theta,\phi)$ according to the semiclassical prescription, are shown \Fig{fig4} for the population imbalance ${A=J_z}$.  In order to evaluate the variance of the fluctuations, we need to calculate the participation number $M$ for a general coherent preparation $|\theta,\pi\rangle$. The result is shown in \Fig{fig4}a. Due to the factorization \Eq{factorization}, this function needs be calculated only once for all desired observables. While we do not have a closed analytic expression for $M(\theta,\phi)$, its characteristic value and its dependence on $u$ and $N$ in the different  phase space regions can be evaluated from general considerations as detailed in Ref.~\cite{Chuchem10}. Generally speaking, the highest participation numbers are obtained at the $\phi=0$ points on the separatrix and scale as 
\be{18}
M \ \ \approx \ \ [\log(N/u)] \ \sqrt{N} ~, 
\eeq
i.e. like the square root of $N$. By contrast, the equatorial coherent states $|\pi/2,0\rangle$ and $|\pi/2,\pi\rangle$ have participation numbers of order unity: 
\be{19}
M(\pi/2,0)&\approx&\sqrt{u}~,
\\ \label{e20} 
M(\pi/2,\pi)&\approx&\sqrt{u}\log(N/u)~.
\eeq

The matrix elements ${(J_z)}_{\nu\mu}$ are shown in \Fig{fig4}b, confirming the assumption of a  broad spectrum containing many frequencies but within a narrow band from the main diagonal. 
The results of the summation over the matrix elements and the integration over the classical fluctuations to obtain the power spectrum $C_A(E)$ according to \Eq{powerspectrum} are compared in  panel \Fig{fig4}c, showing good agreement except for a small region in the vicinity of the separatrix energy.

Similar calculations were carried out for all the Bloch vector components. On the right hand side of \Fig{fig3}c  
we use the participation number $M$ 
and the calculated power spectrum $C_A(E)$ 
for $A=Jz$ in order to predict the variance 
of the population imbalance for the various preparations. 
Similarly for the right hand side of \Fig{fig3}d 
we had to evaluate the total fluctuation intensity 
$C_{\bf S}(E)=\sum_{i=x,y,z} C_{S_i}(E)$ in order to reproduce the combined fluctuations of ${\bf S}$.
Comparison to the results obtained by numerical propagation or by using \Eq{dispers} (left side of the same panels) shows good agreement and confirms the validity of \Eq{factorization}. 

The interpretation of the fluctuation patterns in panels c,d of \Fig{fig3} now becomes clear.  Long time fluctuations will only survive in the vicinity of the unstable equal-population $\phi=\pi$ preparation, where the power spectrum is large and the participation number is small. Note that the classical fluctuations are large for the other separatrix preparations, too, but away from ${\phi=\pi}$ the participation number is large, and hence the quantum fluctuations are quenched. It should also be noted that our approximation quantitatively breaks down in the immediate vicinity of the hyperbolic point $(\theta,\phi)=(\pi/2,\pi)$, because the semiclassical assumption $M\gg 1$ is not satisfied there.

%%%%%%%%%%%%%%%%%%%%%%%%%%%%%%%%%%%%%%%%%%%%%%%%%%%%%%%%%%%%%%%%%%%%%%%%%%%%
\section{Summary}

The magnitude of the long-time quantum fluctuations $\sigma^2_A$
of an arbitrary observable~$A$ can be deduced via \Eq{factorization}
from the tractable classical dynamics, and from the a-priori 
known participation number~$M$ of any coherent preparation $(\theta,\phi)$.
It follows from \Eq{factorization} that the product $M\sigma_A^2$ 
is a function of the energy~$E$ alone. Hence it provided 
essentially the same tomographic information as the average $\overline{A}$.  
But if $\sigma_A^2$ is plotted by itself, its tomographic image
gives complementary and valuable information that goes beyond  
merely mapping energy contours. In particular it allows to detect 
the existence of hyperbolic fixed points and separatrix structures 
which result in distinct participation numbers for the corresponding
coherent preparations.

On the technical level, we see from \Eq{factorization} that $\sigma_A^2$ is 
proportional to the inverse participation number $1/M$, 
hence it has large variation that is implied by \Eq{e18}-\Eq{e20}.
This large variation is illustrated in \Fig{fig4}a,  
and ensures good visibility of the pertinent structures, 
as observed in \Fig{fig3}cd.

It is important to realize that in the classical limit ($N\rightarrow\infty$, 
hence ${M\rightarrow\infty}$) the expectation value 
of any observable relaxes and becomes time-independent, 
with no fluctuations, due to the uniform smearing of the initial coherent 
distribution along the classical trajectories.
{\em Hence the complementary information of $\sigma^2$ tomography is 
available only in a quantum-mechanical reality.} Strangely enough it 
is quantum mechanics that provides an easy way to detect hyperbolic points.
Quantum tomography is a way to probe fine phase space structures that 
would become unsolvable in a classical reality.

%%%%%%%%%%%%%%%%%%%%%%%%%%%%%%%%%%%%%%%%%%%%%%%%%%%%%%%%%%%%%%%%%%%%%%%%%%%%
\section{Acknowledgments}

This research was supported by the Israel Science Foundation (grant Nos. 346/11 and 29/11) and by the United States-Israel Binational Science Foundation (BSF).

%%%%%%%%%%%%%%%%%%%%%%%%%%%%%%%%%%%%%%%%%%%%%%%%%%%%%%%%%%%%%%%%%%%%%%%%%%%

\clearpage
\end{document}